\documentclass[preprint]{iucr}   
\usepackage{tabularx}
\usepackage{amsmath,amsfonts}
\usepackage{algorithm}
\usepackage{algpseudocode}
\usepackage{amssymb}
\usepackage{array}
\usepackage{textcomp}
\usepackage{hyperref}
\usepackage{url}
\usepackage{verbatim}
\usepackage{graphicx}
\usepackage{multirow}
\usepackage{booktabs}
\usepackage{makecell}
\usepackage{paralist}
\usepackage{rotating}
\usepackage{colortbl}
\usepackage{chngcntr}
\usepackage[normalem]{ulem}
\newcommand{\R}{\mathbb{R}}
\DeclareMathOperator*{\argmin}{arg\,min}

\definecolor{lightgray}{gray}{0.9}  

\makeatletter
\newcommand{\mycaption}{%
\ifx \@captype \@undefined \@latex@error {\noexpand \caption outside float}\@ehd \expandafter \@gobble \else \refstepcounter \@captype \expandafter \@firstofone \fi {\@dblarg {\@caption \@captype }}%
}%
\makeatother

     \journalcode{S}              

\begin{document}                  



\title{Multi-stage Deep Learning Artifact Reduction for Parallel-beam Computed Tomography}


\cauthor[a]{Jiayang}{Shi}{j.shi@liacs.leidenuniv.nl}{address if different from \aff}
\author[a]{Dani\"{e}l M.}{Pelt}
\author[a]{K. Joost}{Batenburg}

\aff[a]{Einsteinweg 55, 2333CC, Leiden, \country{the Netherlands}}









\maketitle                        

\begin{abstract}
Computed Tomography (CT) using synchrotron radiation is a powerful technique that, compared to lab-CT techniques, boosts high spatial and temporal resolution while also providing access to a range of contrast-formation mechanisms. The acquired projection data is typically processed by a computational pipeline composed of multiple stages. Artifacts introduced during data acquisition can propagate through the pipeline, and degrade image quality in the reconstructed images. Recently, deep learning has shown significant promise in enhancing image quality for images representing scientific data. This success has driven increasing adoption of deep learning techniques in CT imaging. Various approaches have been proposed to incorporate deep learning into computational pipelines, but each has limitations in addressing artifacts effectively and efficiently in synchrotron CT, either in properly addressing the specific artifacts, or in computational efficiency.

Recognizing these challenges, we introduce a novel method that incorporates separate deep learning models at each stage of the tomography pipeline—projection, sinogram, and reconstruction—to address specific artifacts locally in a data-driven way. Our approach includes \textit{bypass connections} that feed both the outputs from previous stages and raw data to subsequent stages, minimizing the risk of error propagation. Extensive evaluations on both simulated and real-world datasets illustrate that our approach effectively reduces artifacts and outperforms comparison methods. 

\end{abstract}

\section{Introduction}
\begin{figure}
\centering
\includegraphics[width=1.0\textwidth]{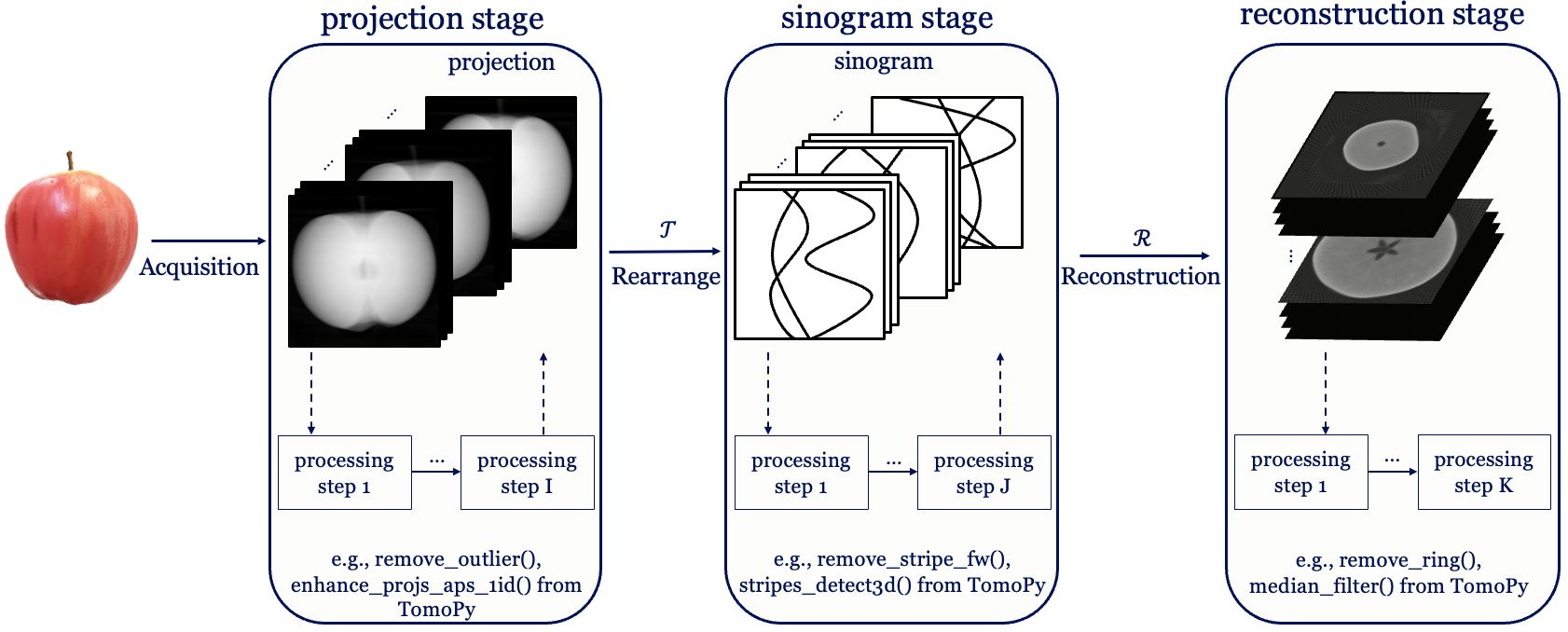}
\caption{Schematic representation of a typical synchrotron CT pipeline, consisting of three stages: projection, sinogram, and reconstruction stage. Each stage may consist of multiple processing steps.}
\label{fig:pipeline}
\end{figure}

X-ray Computed Tomography (CT) is a non-invasive imaging technique widely employed in fields ranging from medical diagnostics to industrial non-destructive testing and security screening \cite{hansen2021computed}. Synchrotron facilities are at the high-end of X-ray tomography capabilities, providing the flux required to image a range of specimens at high spatial and temporal resolutions. \cite{thompson1984computed, westneat2008advances, sun2018synchrotron, labriet2018significant}. Moreover, the synchrotron environment provides access to a range of imaging modes for probing absorption, phase, fluorescence, diffraction and other contrast formation mechanisms. As illustrated in Figure \ref{fig:pipeline}, synchrotron tomography datasets are typically processed by a computational pipeline \cite{hintermuller2010image, gursoy2014tomopy, ganguly2021improving}, which involves three sequential stages:
\begin{inparaenum}[i)] 
    \item acquiring X-ray projection images,
    \item converting these projections into sinogram images,
    \item reconstructing to visualize the object's internal structure.
\end{inparaenum} 
Each stage may consist of multiple processing steps. Implementations of such pipelines can be found in several open-source software packages \cite{gursoy2014tomopy,van2015astra,biguri2016tigre,kazantsev2022high,kim2023differentiable}, such as TomoPy, Astra Toolbox, and LEAP. 

The quality of the reconstructed images is often compromised by various artifacts introduced during data acquisition. These artifacts propagate through the processing pipeline and result in artifacts in reconstructed images. For example, ring artifacts arise due to variations in the detector response and excessively high-energy photons captured by the detector can lead to the occurrence of zinger artifacts. This propagation of artifacts can be mitigated by including artifact-specific data processing steps early in the computational pipeline, such as processing sinogram images for ring artifact reduction \cite{rivers1998tutorial,titarenko2010analytical,paleo2015ring,titarenko2016analytical,vo2018superior,makinen2021ring} and processing projections images for zinger artifact reduction \cite{rivers1998tutorial,gursoy2014tomopy,farago2022tofu,mertens2015method}. While these methods are commonly used in practice, they necessitate parameter tuning to reduce artifacts properly and can introduce additional artifacts, especially if the parameters are not set optimally \cite{pelt2018improving}. We acknowledge that while the effort required for hand-tuning may be relatively modest, significant expertise is often required.
 
\begin{figure}
\centering
\includegraphics[width=0.7\textwidth]{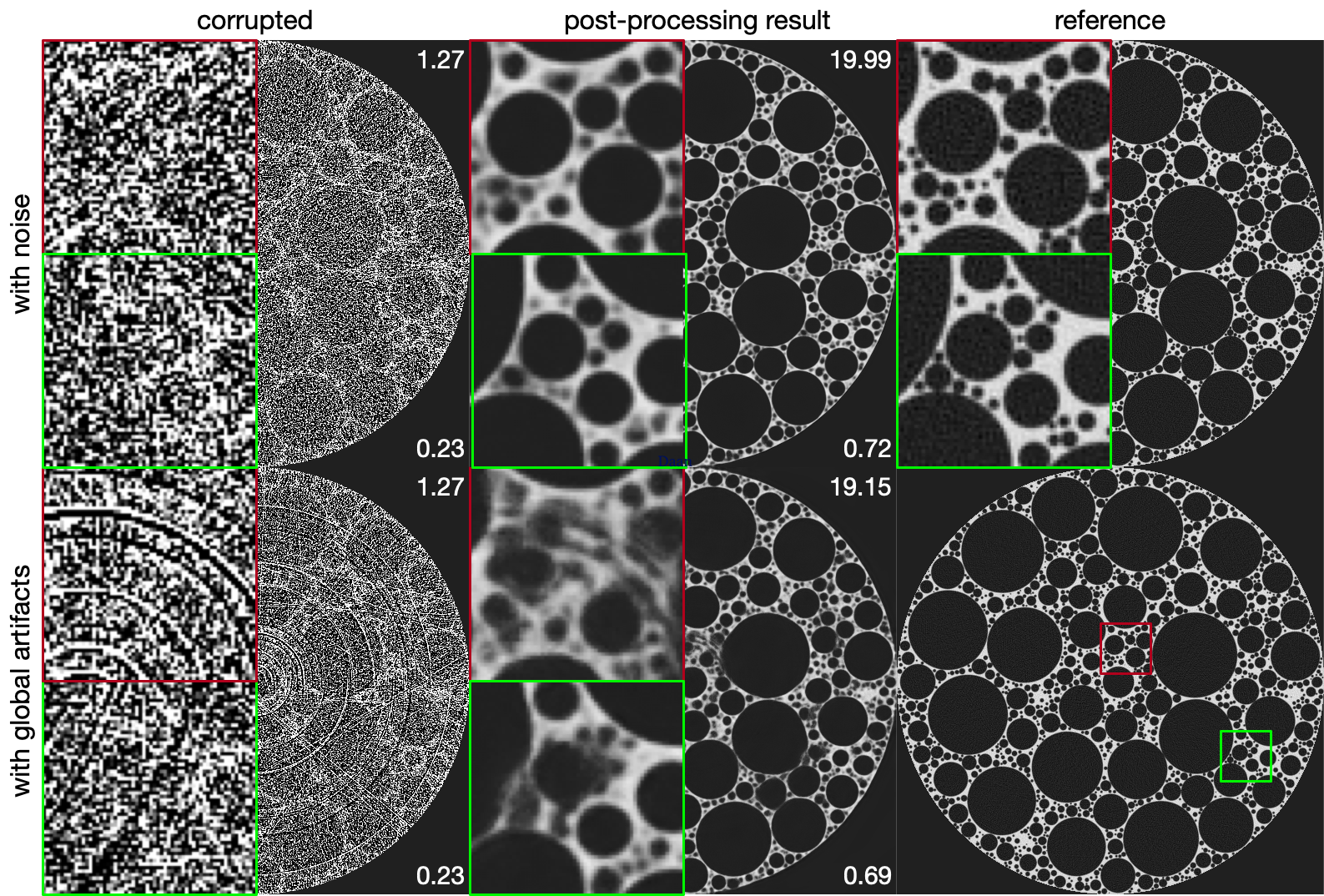}
\caption{Comparison of deep learning-based post-processing performance on reconstructed images from the simulated foam phantom dataset \cite{pelt2022foam} affected by local artifacts (noise) and non-local artifacts. The red and green insets show enlarged views of the affected areas. Peak Signal to Noise Ratio (PSNR) and Structural Similarity Index Measure (SSIM) \cite{wang2004image} values are provided in the top-right and lower-right corners, respectively. Deep learning-based post-processing is \textit{less effective} in reducing global artifacts than noise. While this figure demonstrates differences in artifact reduction based on simulation, similar challenging scenarios can occur in real-world dynamic experiments \cite{buhrer2019high} or fast scan setups \cite{raufaste2015three,mokso2017gigafrost} at synchrotron facilities.}
\label{fig:global_local_artifact}
\end{figure}

Meanwhile, deep learning has significantly advanced the state-of-the-art in image quality enhancement for a variety of imaging tasks, including denoising, super-resolution, and deblurring for natural images \cite{burger2012image, zhang2017beyond, pan2016blind, liang2021swinir}. Its effectiveness has led to increased use in CT imaging \cite{pelt2018improving,nauwynck2020ring,liu2023detector,zhu2023sinogram,shi2023sr4zct}. 
Existing efforts toward integrating deep learning in the synchrotron tomography pipeline roughly falls into three different categories:
\begin{itemize}
    \item \textbf{Integrating deep-learning based postprocessing steps after the image reconstruction step:} Deep learning models, particularly CNNs, are often applied on reconstructed images to enhance image quality \cite{pelt2018improving,chen2017low1,yang2018low,gholizadeh2020deep,yan2023image,liu2023detector}. Although these methods work well for noise reduction, they struggle with non-local artifacts that cannot be easily addressed using local image information, as shown in Figure \ref{fig:global_local_artifact}. Additionally, coupling these models with classical methods that operate before reconstruction introduces the risks associated with parameter tuning and new artifact creation.

    \item \textbf{Replacing individual blocks in the computational pipeline with deep learning modules for specific artifact reduction tasks:} By incorporating deep learning modules into the CT pipeline at the projection and sinogram stages and training them in a supervised manner for a specific artifact-reduction task, classical algorithms for artifact reduction can be replaced by learned methods \cite{yuan2018sipid,ghani2019fast,nauwynck2020ring,fu2023deep,zhu2023sinogram}. However, these models only deal with a specific step of the computational pipeline and may introduce new artifacts in certain cases. When applied in sequence together with post-processing, newly introduced artifacts can propagate through the pipeline and result in artifacts in the reconstructed images.
    
    \item \textbf{Replacing the entire pipeline with end-to-end deep learning:} Ideally, a fully integrated end-to-end deep learning pipeline \cite{wang2020end,chen2022sam} would comprehensively address mentioned challenges such as error propagation, the introduction of new artifacts, and the insufficient reduction of non-local artifacts. However, the vast data size in synchrotron tomography experiments makes this approach computationally impractical, transforming the task into a quasi-3D problem even when using 2D networks. The data volumes typical for synchrotron tomography \textit{far exceed} the memory capacities of most modern GPUs. For example, consider a dataset with a $2000^3$ volume and 2000 projection angles, which demands about 164.8 GB \footnote{The end-to-end workflow includes six stages: data input, projection network processing, sinogram network processing, reconstruction, and final network processing on the reconstruction output. For end-to-end training, it is necessary to host 3D volumes multiple times as input, output, and intermediate data between stages. The memory calculation considers hosting the 3D volume five times to account for these requirements, along with managing three sets of depth-100 2D image features processed by the MS-D network at each stage. The primary memory consumption comes from hosting the $2000^3$ volume five times ($5\times 32.96$ GB), which significantly exceeds the memory required for storing the network’s intermediate features three times ($3\times 1.6$ GB).} of memory just for a single forward inference using a 2D mixed-scale dense (MS-D) network \cite{pelt2018mixed} with a depth of 100. Such large datasets are not only common in synchrotron-based setups but are also increasingly encountered in laboratory-based CT systems, especially when using large flat-panel detectors for high-quality scans with a high number of projections. The primary distinction between these systems lies in the high-throughput acquisition enabled by the larger temporal resolution of synchrotron CT. The memory requirement for \textit{end-to-end training}, which still requires gradients and other computational overhead for back propagation, makes such an approach impractical.
\end{itemize}

The aim of this work is to propose a novel deep learning-based method that is specifically designed for artifact reduction throughout the synchrotron tomography pipeline that combines the ability to address multiple types of artifacts simultaneously, each in their own data domain, with the computational efficiency required for processing large synchrotron tomography datasets. We address the above mentioned challenges by:
\begin{itemize}
    \item \textbf{Comprehensive integration:} We incorporate deep learning across all three CT processing stages—projections, sinograms, and reconstructions—targeting specific artifacts at each stage to exploit deep learning's full potential while simplifying artifact reduction by maintaining their locality.

    \item \textbf{Bypass connections:} We include \textit{bypass connections} that provide each stage access not only to all outputs from preceding stages but also to their raw inputs, reducing the risk of error propagation throughout the pipeline.

    \item \textbf{Efficient training:} To avoid the computational complexity of fully end-to-end training, we adopt 2D CNNs for each stage and propose to train them individually to ensure computation efficiency and practical applicability.
\end{itemize}

This paper is organized as follows. Section \ref{section2} provides an overview of the related concepts and notation that underlie our motivation for the proposed method. In Section \ref{section3}, we describe the details of our method, which involves using a series of CNNs to reduce artifacts at different stages of the processing pipeline, and describe ways of obtaining high-quality reference data for training. Section \ref{section4} covers the experimental design and implementation specifics. In Section \ref{section5}, we present and analyze the experimental results. Section \ref{section6} is dedicated to discussing the implications and significance of our findings. Lastly, in Section \ref{section7}, we conclude the paper by highlighting potential application areas for our method.

\section{Notation and concepts}
\label{section2}
In this section, we first provide a comprehensive overview of the CT pipeline that forms the basis of our method. We then define the general notation for CT artifacts that arise from the acquisition process and show how they propagate through the pipeline, leading to artifacts in the reconstructed images. Next, we discuss deep learning-based denoising methods for CT images and define their notations. Finally, we introduce and define the notations for classical artifact reduction operations that are performed at different steps of the pipeline. This foundation establishes the context and terminology for our proposed artifact reduction method.

\subsection{CT pipeline}
We apply our multi-stage method to a typical synchrotron CT processing pipeline with three sequential steps \cite{hintermuller2010image, gursoy2014tomopy, ganguly2021improving}, illustrated in Figure \ref{fig:pipeline}. First, the CT system scans the object and acquires a series of projection images. Second, those projection images are rearranged into sinogram images. Finally, the reconstructed images are computed.

In X-ray CT, the projection of an object from a specific angle is fundamentally described by the Beer-Lambert law, which models the attenuation of X-rays as they pass through the object. This law states that the intensity $I$ of X-ray radiation exiting the object is exponentially related to the integral of the object's linear attenuation coefficients $\mu(x,y,z)$ along the path $s$ of the beam through the object. Mathematically, this relationship is expressed as $I = I_0 e^{-\int \mu(x,y,z)ds}$, where $I_0$ is the initial intensity of the X-ray beam. We define a projection image as the intensities captured by the detector along a certain angle. Given a detector with $M \times N$ pixels, the object is scanned with $N_{a}$ scanning angles, producing a set of projection images $\mathbf{p} \in \R^{N_{a} \times M \times N}$. Each projection image is influenced by inherent imaging noise such as Poisson noise, arising from the statistical nature of X-ray photon detection. Additionally, artifacts caused by systematic errors in the projection image, like ring artifacts, can be addressed by techniques such as flat-field correction \cite{prell2009comparison,van2015dynamic}.

We first introduce the rearrange operation, denoted as $\mathcal{T}$, where $\mathcal{T}: \R^{N_{a} \times M \times N} \rightarrow \R^{M \times N_{a} \times N}$. We then define sinogram images as the rearranged projection images $\mathbf{s} \in \R^{M \times N_{a} \times N}$. In the reconstruction stage, reconstruction methods are applied to the sinogram images to compute the reconstructed images $\mathbf{r} \in \R^{Z \times Y \times X}$. The reconstruction operation is $\mathcal{R}$, where $\mathcal{R}: \R^{M \times N_{a} \times N} \rightarrow \R^{Z \times Y \times X}$. The projection images $\mathbf{p}$, sinogram images $\mathbf{s}$, and reconstructed images $\mathbf{r}$ represent the same underlying object and are three data representations in the pipeline.

\subsection{Artifacts}
For a general artifact introduced by the imaging process, we illustrate its propagation through the pipeline, leading to artifacts in the reconstructed images. Additionally, we provide schematic representations of the different artifacts in each pipeline stage to facilitate a better understanding of their characteristics.

Corrupted projections $\hat{\mathbf{p}}=\left\{ \hat{\mathbf{p}}_1,...,\hat{\mathbf{p}}_{N_{a}} \right\}$ consist of a series of corrupted projection images $\hat{\mathbf{x}}_i$, and are a combination of the underlying clean projection images $\mathbf{p}$ and artifacts $\mathbf{n}$,
\begin{equation}
\label{equation:noiseinprojection}
\hat{\mathbf{p}} = \mathbf{p} + \mathbf{n}.
\end{equation}
For example, $\mathbf{n}$ can contain noise, offsets by miscalibrated detector pixels that cause ring artifacts, and/or outliers that cause zinger artifacts. In a scenario where no artifact removal steps are included in the CT pipeline and, for illustration, we assume a linear reconstruction operation $\mathcal{R}_{lin}$, the resulting reconstruction images would be expressed as 
$\hat{\mathbf{r}}$:
\begin{equation}
\label{equation:noiseinreconstruction}
\hat{\mathbf{r}} = \mathcal{R}_{lin}\left(\mathcal{T}\left(\hat{\mathbf{p}}\right)\right) = \mathcal{R}_{lin}\left(\mathcal{T}\left(\mathbf{p}\right)\right) + \mathcal{R}_{lin}\left(\mathcal{T}\left(\mathbf{n}\right)\right).
\end{equation}
Here, $\mathcal{R}_{lin}\left(\mathcal{T}\left(\mathbf{p}\right)\right)$ represents the ideal artifact-free reconstruction image, and $\mathcal{R}_{lin}\left(\mathcal{T}\left(\mathbf{n}\right)\right)$ represents the artifacts originating from artifact term $\mathbf{n}$ in corrupted projection images. In this manner, artifacts that occur during acquisition are passed through the pipeline and become artifacts in the reconstructed images. In the following, we describe three common artifact types encountered in (high-energy) CT systems. Existing methods for reducing artifacts are discussed in Section~\ref{sec:classmet}.

\begin{figure}
\centering
\includegraphics[width=0.7\textwidth]{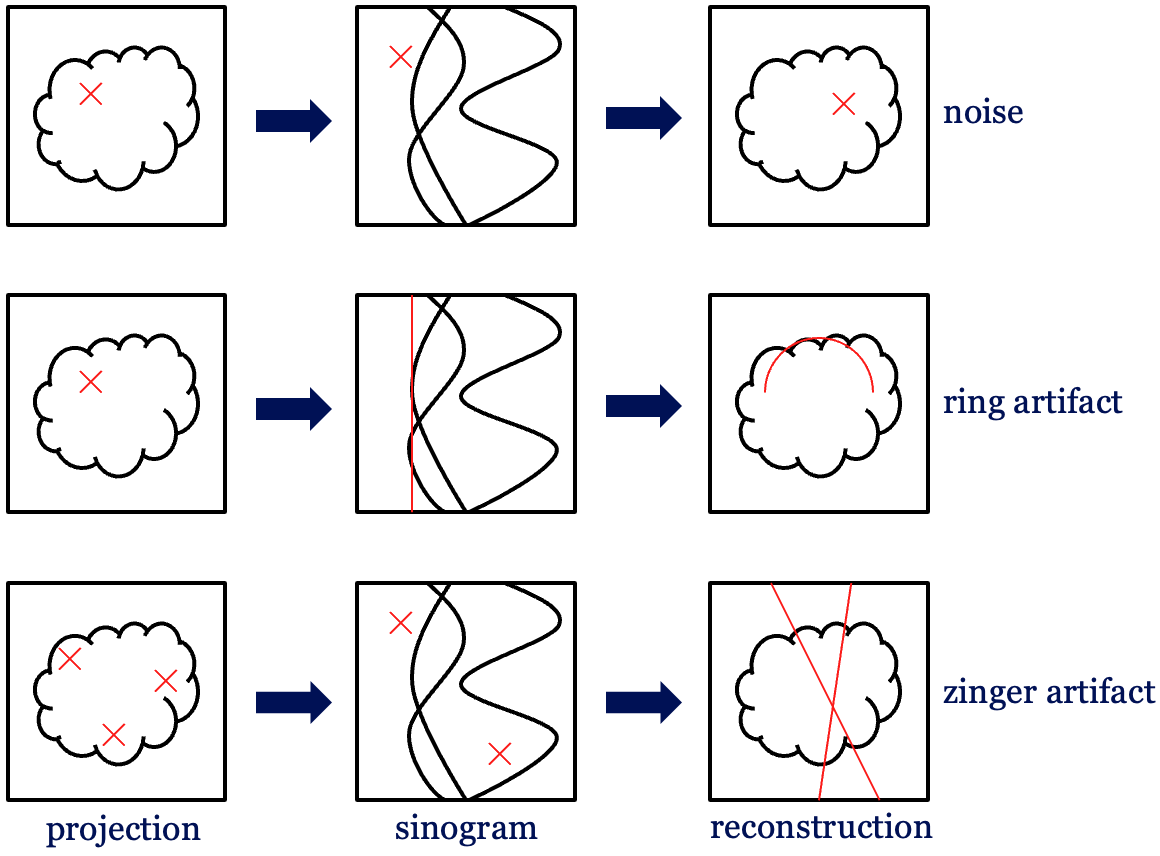}
\caption{Representations of noise, ring, and zinger artifact in projection, sinogram, and reconstruction stages. Red patterns are schematic illustrations of distortions. Noise is a local artifact in images of all stages. Distorted pixel values in the same positions of projection images become a line in the sinogram, resulting in a ring artifact in the reconstructed image. Extremely high pixel values (randomly distributed in different positions) remain as high-value spots in the sinogram and cause crossing lines in the reconstructed image as zinger artifacts.}
\label{artifactsinstage}
\end{figure}

\textbf{Noise:}
Poisson noise is a common artifact in CT, arising from insufficient photon counts at the detector \cite{boas2012ct}. Low-dose CT can suffer from strong noise artifacts due to fewer photons captured by the detector pixels. Poisson noise is introduced during acquisition and presents as local disruptions in projection images. It corresponds to a local perturbation in all stages of the CT pipeline, as illustrated in Figure \ref{artifactsinstage}.

\textbf{Ring Artifact:}
Ring artifacts arise from systematic detector errors, such as miscalibrated or defective elements in the detector. For example, as demonstrated in Figure \ref{artifactsinstage}, a detector element may record its value with an additive offset applied to the actual data \cite{boas2012ct, pelt2018ring}. Consistent detector offsets in the projection images translate to straight lines in sinogram images and become ring-like artifacts in the reconstructed images. 

\textbf{Zinger Artifact:}
Zinger artifacts often appear in high-energy CT, such as synchrotron CT. It is caused by extremely high-value spots in projection images because the detector occasionally records high-energy photons \cite{mertens2015method}. Since these spots' occurrence is stochastic among projections, they appear as prominent local spots in sinograms, as shown in Figure \ref{artifactsinstage}. After reconstruction, the local artifacts become crossing streaks in the reconstructed images.  


\subsection{Post-processing with Deep Learning }
\label{sec:postprocessing}
We denote the CNN as $f_\theta$, with $\theta$ representing the network's learnable parameters. In the context of CT image post-processing, the CNN operates on a corrupted reconstruction image $\hat{\mathbf{r}}$ to yield a processed output $\mathbf{r}^{\text{PP}}$ as follows:
\begin{equation}
\label{eq:postprocess}
\mathbf{r}^{\text{PP}} = f_\theta(\hat{\mathbf{r}})
\end{equation}
This paper focuses on 2D CNNs for their \textit{computational efficiency} over 3D counterparts, especially important given the large size of CT images, which often exceed $1000^3$ pixels. Thus, $f_\theta$ is applied slice-by-slice across the $Z$ slices of $\hat{\mathbf{r}}$, aggregating the outputs into $\mathbf{r}^{\text{PP}}$.

To optimize the CNN parameters $\theta$ to approximate artifact-free reconstructions, we employ supervised learning with a training set $X = {(\hat{\mathbf{r}}_1, \mathbf{r}_1^{\text{HQ}}), \ldots, (\hat{\mathbf{r}}_{N^t}, \mathbf{r}_{N^t}^{\text{HQ}})}$, pairing $N^t$ corrupted reconstructions $\hat{\mathbf{r}}_i$ with high-quality references $\mathbf{r}_i^{\text{HQ}}$. High-quality reconstructions are typically acquired through high-dose scans, a large number of projection images, or advanced reconstruction techniques \cite{mohan2014model,kazantsev2017novel}. However, consistently acquiring high-quality reconstructions is often impractical in many scenarios due to their high resource demands, long acquisition times, or high radiation exposure. By leveraging CNN-based post-processing, high-quality images can be approximated from low-dose or under-sampled acquisitions, providing an efficient alternative to traditional approaches. This capability is particularly advantageous for data from dynamic experiments \cite{sieverts2022unraveling} or batch processing of similar objects, where rapid and reliable artifact reduction is crucial.

The optimal parameters $\theta^*$ are determined by minimizing the loss function $L$ that quantifies the discrepancy between the CNN outputs and reference images:
\begin{equation}
\label{equationn:deeplearningobjective}
\theta^* = \argmin_{\theta} \sum_{i=1}^{N^t} L\left(f_{\theta}(\hat{\mathbf{r}}_i),\mathbf{r}^{\text{HQ}}_i\right),
\end{equation}.

CNN-based post-processing has been shown to effectively reduce noise across various CT imaging applications. However, CNNs typically learn to exploit local information due to their use of small convolution kernels, even when their depth allows for large receptive fields. Due to their reliance on local convolution kernels, CNNs are generally better suited for mitigating local artifacts than non-local ones, such as ring or zinger artifacts. This limitation is illustrated in Figure \ref{fig:global_local_artifact}, where deep learning post-processing shows notable performance in noise reduction but is less effective against non-local artifacts. This observation underscores the rationale for the multi-stage method proposed in this manuscript.

\subsection{Classical Artifact Reduction Methods}
\label{sec:classmet}
In practice, classical (i.e. non-learning) artifact reduction methods are typically performed at different stages of the CT pipeline. Specifically, ring artifacts are primarily addressed at the sinogram stage, zinger artifacts are tackled within projection images, and denoising techniques are applied to the reconstructed images. We define the following artifact reduction in classical CT imaging workflows as specific operations: \begin{inparaenum}[i)]
\item \textbf{Operation on projection images $\mathbf{A}_p$},
\item \textbf{Operation on sinogram images  $\mathbf{A}_s$},
\item \textbf{Operation on reconstructed images $\mathbf{A}_r$}.
\end{inparaenum}

The integration of these artifact reduction methods into the CT pipeline can be represented by the following formula, which sequentially applies projection stage operator $\mathbf{A}_p$, sinogram stage operator $\mathbf{A}_s$, and reconstruction stage operator $\mathbf{A}_r$ to produce final, artifact-reduced images:
\begin{equation}
\label{equation:artifactreductionobjective}
\mathbf{r}^{\text{PL}} = \mathbf{A}_r\left(\mathcal{R}\left(\mathbf{A}_s\left(\mathcal{T}\left(\mathbf{A}_p\left(\hat{\mathbf{p}}\right)\right)\right)\right)\right).
\end{equation}


In detail, classical denoising methods, such as median or Wiener filtering and TV-based regularization \cite{rudin1992nonlinear}, effectively reduce noise by enhancing image smoothness or exploiting patch similarity (e.g., BM3D \cite{dabov2007image}). These are denoted by $\mathbf{A}_r$ when applied to reconstructed images. Additionally, noise reduction can also be implicitly integrated within the reconstruction algorithm $\mathcal{R}$.

For ring artifacts, strategies often involve filtering techniques applied directly to sinograms to address linear disturbances \cite{rivers1998tutorial,anas2010removal,ketcham2006new, munch2009stripe} or, alternatively, transforming reconstructed images to polar coordinates for line-based artifact correction \cite{sijbers2004reduction,brun2009improved,chao2019removal}. These are captured by $\mathbf{A}_s$ and $\mathbf{A}_r$, respectively. Meanwhile, zinger artifact reduction in projection images $\mathbf{A}_p$ is frequently addressed with filtering methods, such as those provided by software like Tomopy \cite{gursoy2014tomopy}, which includes specialized functions for this purpose.

\section{Algorithm}
\label{section3}

This section outlines our multi-stage artifact reduction approach. We delve into the methodology, detailing the CNN training process and the acquisition of high-quality reference data for training purposes. The discussion concludes by highlighting the design choices that enhance the computational efficiency of our method.

\subsection{Multi-stage Artifact Reduction}
\begin{figure}
\centering
\includegraphics[width=0.7\textwidth]{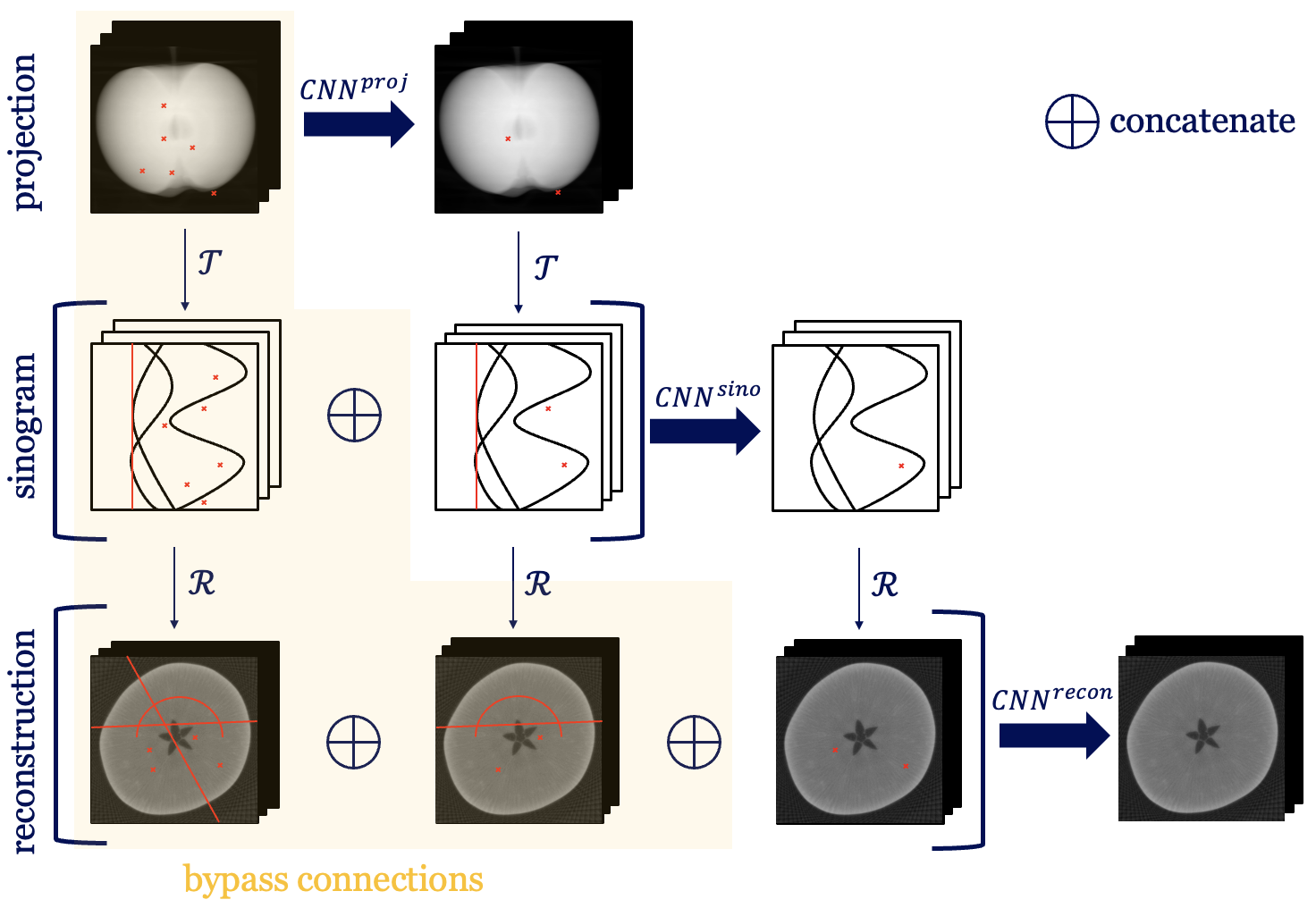}
\caption{Schematic of our proposed multi-stage artifact reduction method, illustrating distortions and their correction at each stage. Our method includes \textit{bypass connections} that incorporate both raw and previously processed data at each stage to reduce the risk of error propagation throughout the pipeline.}
\label{fig:algorithm}
\end{figure}

Our data-driven approach, depicted in Figure \ref{fig:algorithm}, utilizes three CNNs $f_{\theta_p}^{p}$, $f_{\theta_s}^{s}$, and $f_{\theta_r}^{r}$ to sequentially process projection, sinogram, and reconstruction data. To enhance the efficacy of this sequence and reduce the risk of error propagation, our model integrates \textit{bypass connections} that incorporate both raw and previously processed data at each stage. This design not only improves the robustness of artifact reduction across the pipeline but also maintains the integrity of the original data. Each CNN is trained independently in a sequential manner to ensure training efficiency.

Three CNNs are employed independently at each stage of the pipeline, with their specific inputs and outputs defined as follows:
\begin{itemize}
    \item \textbf{Projection stage:} The input to the first network, $f_{\theta_p}^{p}$, is the set of raw projection images $\hat{\mathbf{p}}=\left\{ \hat{\mathbf{p}}_1,...,\hat{\mathbf{p}}_{N_{a}} \right\}$, where $N_{a}$ denotes the number of projections. The output is the set of enhanced projection images $\mathbf{p}^*=\left\{\mathbf{p}_1^*,...,\mathbf{p}_{N_{a}}^* \right\}$, computed as $\mathbf{p}^{*} = f_{\theta_p}^{p}(\hat{\mathbf{p}})$. 
    
    \item \textbf{Sinogram stage:} The second network, $f_{\theta_s}^{s}$, processes the sinograms, which are obtained by rearranging both the raw and enhanced projections using a transformation $\mathcal{T}$. The input to this network consists of both the raw sinograms $\mathcal{T}(\hat{\mathbf{p}})$ and the rearranged enhanced projection images $\mathcal{T}(\mathbf{p}^{*})$. The output is the enhanced sinograms $\mathbf{s}^{*} = f_{\theta_s}^{s}\left(\mathcal{T}\left(\hat{\mathbf{p}}\right), \mathcal{T}\left(\mathbf{p}^{*}\right)\right)$. 

    \item \textbf{Reconstruction stage:} The final network, $f_{\theta_r}^{r}$, refines the reconstructed images by taking as input the reconstructions of all sinograms, including those derived from raw data and the enhanced outputs of previous stages. The input is composed of the reconstructed raw sinograms $\mathcal{R}(\mathcal{T}(\hat{\mathbf{p}}))$, the reconstructed rearranged enhanced projection images $\mathcal{R}(\mathcal{T}(\mathbf{p}^{*}))$, and the reconstructed enhanced sinograms $\mathcal{R}(\mathbf{s}^{*})$. The output is the set of enhanced reconstructed images $\mathbf{r}^{*}$, calculated as $\mathbf{r}^{*} = f_{\theta_r}^{r}\left(\mathcal{R}\left(\mathcal{T}\left(\hat{\mathbf{p}}\right)\right), \mathcal{R}\left(\mathcal{T}\left(\mathbf{p}^{*}\right)\right), \mathcal{R}\left(\mathbf{s}^{*}\right)\right)$. 
\end{itemize}

In essence, this process underscores the strategic application of 2D CNNs across different stages of data processing, emphasizing computational efficiency and systematic enhancement. For example, $f_{\theta_p}^p$ is applied to each of the $N_a$ projection images $\hat{\mathbf{p}}$, culminating in the collection of processed projections $\mathbf{p}^{*}$. This stepwise approach is mirrored in subsequent stages, ensuring a comprehensive and effective artifact reduction through the entire imaging pipeline. The method is summarized in Algorithm \ref{algo:inferenceinpseudocode}.

\begin{algorithm}
\hspace*{\algorithmicindent} Input: low-quality projection images $\hat{\mathbf{p}}$, reconstruction method $\mathcal{R}$, trained networks $f^p_{\theta_p}$, $f^s_{\theta_s}$, and $f^r_{\theta_r}$. \\ 
 \hspace*{\algorithmicindent} Output: Enhanced reconstructions with reduced artifacts $\mathbf{r}^{*}$
 \label{algo:inference}
\begin{algorithmic}[1]
\Procedure{Inference}{} 
\State Process $\hat{\mathbf{p}}$ through $f^p_{\theta_p}$ to get $\mathbf{p}^{*}$.
\State Rearrange $\hat{\mathbf{p}}$ and $\mathbf{p}^{*}$ to sinograms  $\mathcal{T}\left(\hat{\mathbf{p}}\right)$ and $\mathcal{T}\left(\mathbf{p}^{*}\right)$. 
\State Apply $f_{\theta_s}^{s}$ to the two sets of sinograms to obtain output sinograms $\mathbf{s}^{*}$.
\State Reconstruct all data to obtain $\mathcal{R}\left(\mathcal{T}\left(\hat{\mathbf{p}}\right)\right)$, $\mathcal{R}\left(\mathcal{T}\left(\mathbf{p}^{*}\right)\right)$, $\mathcal{R}\left(\mathbf{s}^{*}\right)$
\State Refine reconstructions via $f_{\theta_r}^{r}$ to achieve $\mathbf{r}^{*}$.
\EndProcedure
\end{algorithmic}
\mycaption{Inference method}\label{algo:inferenceinpseudocode}
\end{algorithm}

\subsection{Training Procedure}
Our training procedure, as detailed in Algorithm~\ref{algo:traininpseudocode}, strategically decomposes the multi-stage artifact reduction process into separate segments, optimizing the parameters $\Theta = \{\theta_p, \theta_s, \theta_r\}$ of each CNN independently for \textit{efficiency} and \textit{practicality}. Although our approach employs 2D CNNs for each phase, the incorporation of operations $\mathcal{T}$ and $\mathcal{R}$, along with the inherently 3D nature of CT images, imbues our method with a 3D-like capability for artifact reduction. By avoiding the computationally intensive demands of end-to-end 3D training, our strategy greatly reduces the significant computational demands typically faced by 3D deep learning models. As demonstrated in Section \ref{sec:ablation}, chaining the projection, sinogram, and reconstruction stages outperforms models trained on single stages alone, with further improvements achieved through our proposed \textit{bypass connections}. This approach not only maintains the model's applicability to large-scale CT problems but also ensures that each stage is fine-tuned to its specific target of artifact reduction.

For our training procedure, we primarily employ a supervised learning approach, though alternatives like self-supervised learning are also feasible. Supervised training necessitates high-quality reference projections, sinograms, and reconstructions. These references are attainable through several methods, tailored to the specific requirements of the use case. High-quality scans may involve using increased radiation doses and capturing a large number of projections, from which reference reconstructions are derived. When the projection angles in low-quality scans match those in high-quality scans, corresponding high-quality projections can be directly selected. Alternatively, for non-matching angles, simulations of projections from high-quality reconstructions can be performed, for instance, utilizing tools like the ASTRA toolbox~\cite{van2015astra}. Another approach involves generating artifact-reduced reconstructions through advanced scanning techniques or processing methods designed to minimize artifacts~\cite{zhu2013micro, pelt2018ring}. These reconstructed images can then serve as the basis for simulating high-quality projections and sinograms, ensuring a comprehensive set of reference data for training.

The training begins with optimizing $\theta_p$ for the projection domain, using high-quality reference projections to establish a training set. The objective function is minimized to find optimal parameters. Following this, we proceed to the sinogram domain, optimizing $\theta_s$ based on both corrupted and enhanced sinogram pairs, accommodating for discrepancies in projection angles through upsampling. The process culminates in the reconstruction domain, refining $\theta_r$ by leveraging the outputs of the previous stages and high-quality reference reconstructions to guide the training. Throughout these stages, domain-specific loss functions can be employed to precisely target and mitigate artifacts, ensuring an effective reduction across each domain.

\begin{algorithm}
\hspace*{\algorithmicindent} Input: low-quality projection images $\hat{\mathbf{p}}_i$, high-quality projections $\mathbf{p}^\text{HQ}_i$, sinograms $\mathbf{s}^\text{HQ}_i$, and reconstructions $\mathbf{r}^\text{HQ}_i$ for $N^t$ reference objects. Loss functions $L^p$, $L^s$, and $L^r$.\\
\hspace*{\algorithmicindent} Output: optimized CNN weights: $\theta_p^*$, $\theta_s^*$, $\theta_r^*$.
\begin{algorithmic}[1]
\Procedure{Training}{}
\State Optimize $\theta_p$ in the projection domain: 
\Statex $\hspace{\algorithmicindent}\hspace{\algorithmicindent}\theta_p^* = \argmin_{\theta_p} \sum_{i=1}^{N^t} L^p\left(f_{\theta_p}^p(\hat{\mathbf{p}}_i),\mathbf{p}^{\text{HQ}}_i\right)$.
\State Generate enhanced projections $\mathbf{p}^*_i$ using $f_{\theta_p^*}^p$.
\State Optimize $\theta_s$ in the sinogram domain: 
\Statex $\hspace{\algorithmicindent}\hspace{\algorithmicindent}\theta_s^* = \argmin_{\theta_s} \sum_{i=1}^{N^t} L^s\left(f_{\theta_s}^s(\hat{\mathbf{s}}_i, \mathcal{T}(\mathbf{p}^*_i)),\mathbf{s}^{\text{HQ}}_i\right)$.
\State Produce refined sinograms $\mathbf{s}^*_i$ with $f_{\theta_s^*}^s$.
\State Finalize $\theta_r$ in the reconstruction domain: 
\Statex $\hspace{\algorithmicindent}\hspace{\algorithmicindent}\theta_r^* = \argmin_{\theta_r} \sum_{i=1}^{N^t} L^r\left(f_{\theta_r}^r(\hat{\mathbf{r}}_i, \mathcal{R}(\mathcal{T}(\mathbf{p}^*_i)), \mathcal{R}(\mathbf{s}^*_i)),\mathbf{r}^{\text{HQ}}_i\right)$.
\EndProcedure
\end{algorithmic}
\mycaption{Training procedure}\label{algo:traininpseudocode}
\end{algorithm}

\section{Experiments}
\label{section4}
In this section, we detail the artifact simulation on the simulated dataset and introduce the real-world dataset for evaluating our method. We also cover the implementation details and the metrics used to evaluate our experiments.

\subsection{Datasets}
Our method was assessed using the following datasets:
\begin{itemize}
    \item \textbf{Dataset with Simulated Artifacts} 
    \begin{itemize} \item  \textbf{Foam Phantom}: Using the {\tt foam\_ct\_phantom} package \cite{pelt2022foam}, we created simulated cylinder foam phantoms of dimensions $512\times512\times512$. For each experiment, two unique phantoms were generated for training and testing, respectively. These phantoms contained 100,000 non-overlapping bubbles of various sizes within a cylindrical volume. Projection images were generated at a resolution of $512 \times 512$ pixels across 1024 angles spanning 180 degrees using parallel beam geometry.
    
    \item \textbf{LoDoInd}: From the LoDoInd dataset \cite{shi2024LoDoInd}, we selected the reference tube sample, characterized by its composition of 15 different materials, such as coriander and pine nuts. This material diversity in the LoDoInd dataset presents a higher level of complexity compared to the single-material Foam Phantom dataset, offering a more challenging scenario that effectively evaluates our method's performance on intricate data. The sample's dimensions are $4000\times1250\times1250$ pixels, with the top half designated for training and the bottom for testing. The upper and lower halves of the sample are notably distinct in material composition and structural features, ensuring a meaningful evaluation of the model’s generalization capability. Projection images were simulated over 1024 angles distributed over a 180-degree range, similar to the Foam Phantom setup.
    \end{itemize}
    
    \item \textbf{Dataset with Real-world Artifacts} 
    \begin{itemize} \item \textbf{Tomobank}: The \texttt{fatigue-corrosion} experimental dataset from TomoBank \cite{de2018tomobank} provided 25 distinct tomographic sets of aluminum alloy subjected to fatigue testing. Each set consists of 1500 projection images with dimensions of 2560 × 2160 pixels, uniformly sampled over 180 degrees. Dark field and flat field images were also captured. We utilized \texttt{tomo\_00056} and \texttt{tomo\_00055} for training and testing (before and after the alloy breaks into two pieces in the corrosion experiment, images are fairly different for training and testing), respectively, with volumes sized $2160\times2560\times2560$ pixels.
    \end{itemize}
\end{itemize}

\subsection{Artifact Simulation}
\begin{figure}
\centering
\includegraphics[width=0.7\textwidth]{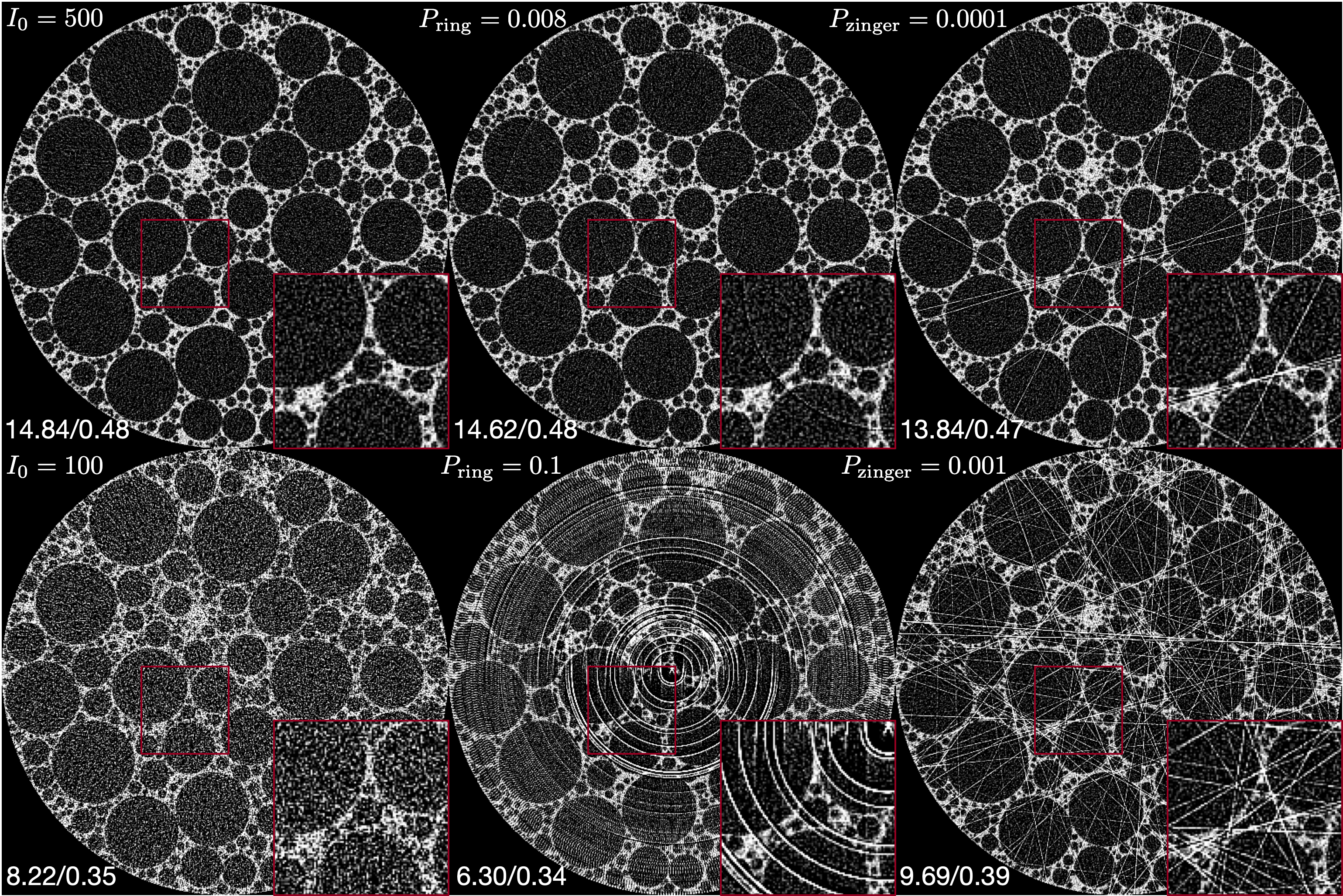}
\caption{Example image with various levels of artifacts on a simulated foam phantom dataset. The artifacts, including noise, ring artifacts, and zinger artifacts, were generated by varying the parameters $I_0$, $P_{\text{ring}}$, and $P_{\text{zinger}}$, respectively. The PSNR and SSIM metrics with respect to the ground truth image are provided for each reconstructed image, displayed in the bottom left.}
\label{compareartifact}
\end{figure}

We applied simulations of noise, ring, and zinger artifacts to the Foam Phantom and LoDoInd datasets. Figure \ref{compareartifact} displays the simulated foam phantom with varying levels of these artifacts.

\textbf{Noise:} 
Low-dose projections were simulated by applying Poisson noise to the projection images, converting the data into raw photon counts following the procedure in \cite{pelt2022foam}. Noise levels were regulated by two factors: the average absorption $\gamma$, set to absorb about half the photons, and the incident photon count $I_0$, which was varied to simulate different noise levels. For low-quality reconstructions, 256 projection images were used, and for high-quality reconstructions, we used 1024 projection images.

\textbf{Ring Artifact:}
Ring artifacts were simulated by introducing fixed errors to randomly selected detector pixels to emulate systematic detector errors. This was quantified by $\mathbf{d}_{\text{ring}} \in \R^{M \times N}$, influenced by the affected pixel percentage $P{\text{ring}}$ and the deviations' standard deviation $\sigma_{\text{ring}}$:
\begin{equation}
\mathbf{d}_{\text{ring}} = \mathbf{M}\left(P_{\text{ring}}\right) \mathcal{N}\left(\mathbf{0},\sigma_{\text{ring}}^{2}\mathbf{I}\right),
\end{equation}
where $\mathbf{M}\left(P_{\text{ring}}\right)$ is a mask with $P_{\text{ring}}$ percent of pixels set to one. Corrupted projections $\hat{\mathbf{p}}$ were obtained by adding $\mathbf{d}_{\text{ring}}$ to each image $\mathbf{x}_i$:
\begin{equation}
\hat{\mathbf{x}}_i = \mathbf{x}_i + \mathbf{d}_{\text{ring}}.
\end{equation}
The artifact's severity is modulated by $P_{\text{ring}}$, with $\sigma_{\text{ring}}$ fixed at 0.005.

\textbf{Zinger Artifact:}
Zinger artifacts were simulated by setting random pixel values in projection images to excessively high levels, imitating detector saturation. The impact of zingers was determined by the percentage of affected projections $P_{\text{proj}}$, set at 10\%, and the percentage of altered pixels within these projections $P_{\text{zinger}}$. The excessive value $v$ was set at 5, with the variation in $P_{\text{zinger}}$ affecting the number of streaks in the reconstructed slices.

\subsection{Preparation of Tomobank Dataset}
For the Tomobank dataset, we processed the data to generate both high-quality and corresponding low-quality versions. High-quality data were refined through flat-field correction, utilizing the median values from all ten available flat and dark fields. Despite these initial corrections, slight ring and zinger artifacts remained in the reconstructed images. To mitigate these, we employed TomoPy \cite{gursoy2014tomopy} for further refinement of high-quality data: applying a median filter to remove zinger artifacts from projection images and utilizing a polar coordinate system for ring artifact removal in the reconstructed images \cite{sijbers2004reduction}. The parameters for these additional steps were determined through visual inspection. For low-quality data creation, we selected a single flat field and dark field at random for correction and reduced the set to 500 equally-spaced projection images from the original 1500. No additional zinger or ring removal was conducted for low-quality data.

\subsection{Comparison Methods}
Our method's effectiveness was benchmarked against various established approaches to validate its performance in artifact reduction:
\begin{itemize}
    \item \textbf{Post-proc.}: Involves deep learning-based post-processing techniques applied directly to reconstructed images to enhance their quality.
    
    \item \textbf{Sinogram proc.}: A deep learning approach for processing sinogram data before reconstruction, as introduced by \cite{nauwynck2020ring}, which employs a line suppression loss function to address artifacts.

    \item \textbf{MBIR/Kazantsev}: Employs a model-based regularized iterative reconstruction (MBIR) method that incorporates a data fitting term based on the Student's t-distribution \footnote{https://github.com/dkazanc/ToMoBAR}, as proposed by \cite{kazantsev2017novel}. This method aims to suppress outliers and enhance reconstruction from limited data, complemented by subsequent deep learning-based post-processing.
    
    \item \textbf{Median filter + M{\"u}nch}: Utilizes a ring artifact removal technique based on Wavelet-Fourier filtering, suggested by \cite{munch2009stripe}, and is paired with median filtering to specifically target zinger artifacts. Deep learning-based post-processing is applied after initial pre-processing and reconstruction to further refine image quality.
    
    \item \textbf{Median filter + Miqueles}: Applies a generalized Titarenko's algorithm for ring artifact reduction, as developed by \cite{miqueles2014generalized}, and integrates median filtering. Following pre-processing and reconstruction, deep learning-based post-processing is employed to enhance the final image quality.
    
    \item \textbf{Vo}: Implements a ring reduction algorithm by \cite{vo2018superior}, designed to address various types of striping artifacts in sinograms. This method includes a median filtering step, making it effective against both ring and zinger artifacts. Similarly, deep learning-based post-processing is adopted after pre-processing and reconstruction.
\end{itemize}

For classical methods, we \textit{optimized parameters} through grid search using the training dataset. We iterated over parameter combinations, computing reconstructions for each and comparing them against reference reconstructions free of artifacts from the training dataset to identify the most effective parameter set by choosing the highest SSIM value among them. This optimization process was tailored for each dataset and level of simulated artifact.

\subsection{Implementation Details}
In this study, we employed the MS-D network \footnote{https://github.com/ahendriksen/msd\_pytorch} \cite{pelt2018mixed} for all three stages of our method, setting the network depth to 100. The total number of trainable parameters for each stage is detailed in Table \ref{trainableparameters}. To ensure a fair comparison with deep learning-based post-processing methods, we adjusted the MS-D network's depth to 180 in those cases, aligning the parameter count across different approaches, as Table~\ref{trainableparameters} illustrates.

\begin{table}
  \centering
\caption{Number of trainable parameters of used neural networks in our proposed method (total) and the deep learning-based post-processing method (post-proc.).}
\begin{center}
\label{trainableparameters}
    \begin{tabular}{ccccc|c} \toprule[0.1pt]
     network & stage 1 & stage 2 & stage 3 & total & post-proc.  \\ \midrule[0.05pt]
        MS-D & 45652 & 46553 & 47454 & 139659 & 146972\\ \midrule[0.05pt]
    \end{tabular}
\end{center}
\end{table}

Training of the networks occurred over 200, 200, and 500 epochs for the first, second, and third stages, respectively, using the ADAM optimizer \cite{kingma2014adam} with an L2 loss function and an initial learning rate of $10^{-3}$. The training for the final stage is longer as there are more complicated image features in reconstructed images. Comparable total training epochs (900) were assigned for the post-processing networks to match the combined duration. Training was subject to early stopping, triggered by 10 epochs without validation loss improvement or exceeding 14 days.

Performance evaluation utilized Peak Signal to Noise Ratio (PSNR) and Structural Similarity Index Measure (SSIM) \cite{wang2004image}, computed against high-quality reference data based on the reference data's range.

\section{Results}
\label{section5}

\subsection{Artifact Reduction Performance}
\label{sec:section5_artifact_reduction_performance}
\begin{figure}
\centering
\includegraphics[width=0.85\textwidth]{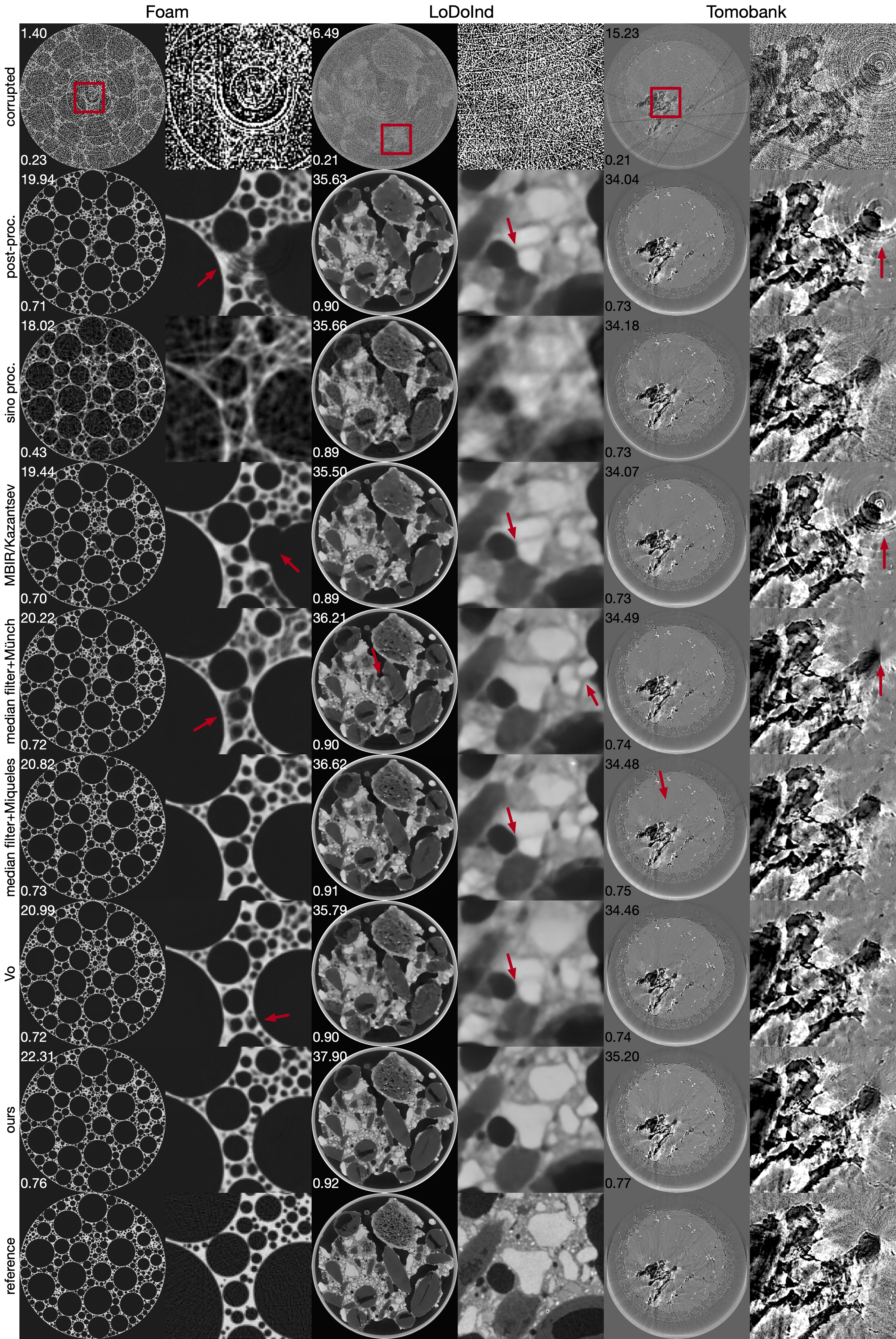}
\caption{Comparative visualization on three different datasets of our proposed artifact reduction method against various established techniques, encompassing strategies that initially apply classical pre-processing followed by deep learning-based post-processing, as well as methods employing deep learning-based processing across distinct stages. Red insets highlight magnified sections for closer inspection. Red arrows point out the remaining artifacts.}
\label{fig:compare_artifact_reduction}
\end{figure}

Table \ref{tab:artifact_reduction_comparison} presents a quantitative comparison between our method and other artifact reduction approaches. Our method consistently surpasses all others in PSNRs and SSIMs across datasets with both simulated and real-world artifacts. It was observed that classical artifact reduction methods, when followed by deep learning-based post-processing, generally achieved better PSNRs and SSIMs than strategies that rely solely on post-processing. This highlights the utility of classical artifact reduction techniques. Among the evaluated methods, those focusing exclusively on sinogram images were the least effective in compared metrics, primarily because they omit the crucial image enhancement step applied to reconstructed images. Figure \ref{fig:compare_artifact_reduction} visually corroborates these findings: our approach not only delivers superior image clarity but also more effectively addresses areas heavily affected by global artifacts, unlike the competing methods which frequently struggle in such regions.

\begin{table}
\caption{Evaluative comparison across deep learning-based, classical, and our proposed methods on the simulated foam phantom \cite{pelt2022foam}, LoDoInd \cite{shi2024LoDoInd} and real-world experimental dataset Tomobank \cite{de2018tomobank}. Performance metrics, specifically PSNR and SSIM, are averaged across all slices and highlighted in bold for the best outcomes. Classical methods include an additional post-processing step before reconstruction, with parameters optimized on training data. All methods were designed to have similar numbers of trainable parameters for the same network architecture and were trained for a consistent number of epochs to ensure fairness in comparison.}
\label{tab:artifact_reduction_comparison}

\centering
\resizebox{\textwidth}{!}{
\setlength{\tabcolsep}{2pt}
\begin{tabular}{ccccccccc}  \toprule[0.1pt]
  Parameters & \multirow{2}{*}{corrupted} & \multirow{2}{*}{post-proc.} & \multirow{2}{*}{sinogram proc.} &  MBIR/Kazantsev & M{\"u}nch &  Miqueles &  Vo &  \multirow{2}{*}{ours} \\ \cmidrule{5-8}
 $I_0/P_{\text{ring}}/P_{\text{zinger}}$ & & & & \multicolumn{4}{c}{followed by post-proc.} & \\ \midrule[0.05pt]  

  \multicolumn{9}{c}{\cellcolor{lightgray} Dataset: Foam ($512,512,512$)} \\ \midrule[0.05pt]  

 30/0.1/0.001 & 1.14/0.23 & 19.55/0.70 & 17.97/0.44 & 19.24/0.70 & 19.72/0.71 & 20.34/0.72 & 20.37/0.71  & \textbf{21.80}/\textbf{0.76}  \\ \cmidrule{2-9} 

 100/0.1/0.001 & 4.07/0.27 & 21.10/0.69 & 19.58/0.51 & 21.81/0.75 & 22.26/0.76 & 23.70/0.77 & 23.57/0.78  & \textbf{24.24}/\textbf{0.79}  \\ \cmidrule{2-9}

 100/0.1/0 & 4.97/0.29 & 22.67/0.77 & 19.63/0.51 & 23.00/0.77 & 22.39/0.76 & 23.32/0.77 & 23.38/0.77  & \textbf{24.41}/\textbf{0.79}  \\ \cmidrule{2-9}

 100/0.2/0 & 3.04/0.26 & 22.21/0.76 & 19.59/0.51 & 22.50/0.76 & 22.25/0.75 & 22.78/0.76 & 23.20/0.77  & \textbf{24.09}/\textbf{0.78}  \\ \cmidrule{2-9}

 100/0/0.002 & 5.58/0.29 & 23.10/0.77 & 19.66/0.51 & - & 24.26/0.78 & - & 23.86/0.77  &  \textbf{24.87}/\textbf{0.79}  \\ \midrule[0.05pt]

 \multicolumn{9}{c}{\cellcolor{lightgray} Dataset: LoDoInd ($2000,1250,1250$)} \\ \midrule[0.05pt] 

 500/0.1/0.001 & 5.91/0.21 & 36.30/0.91 & 36.22/0.90 & 36.11/0.91 & 36.93/0.92 & 36.93/0.92 & 36.50/0.91  & \textbf{38.65}/\textbf{0.93} 

\\ \midrule[0.05pt] 

\multicolumn{9}{c}{\cellcolor{lightgray} Dataset: Tomobank ($2160,2560,2560$)} \\ \midrule[0.05pt] 

 NA & 17.77/0.24 & 35.64/0.77 & 35.33/0.77 & 35.53/0.77 & 35.77/0.78 & 35.79/0.78 & 35.81/0.78  & \textbf{36.55}/\textbf{0.79} 

\\ 
\bottomrule[0.1pt]
\end{tabular}
}
\end{table}

\begin{figure}
\centering
\includegraphics[width=1\textwidth]{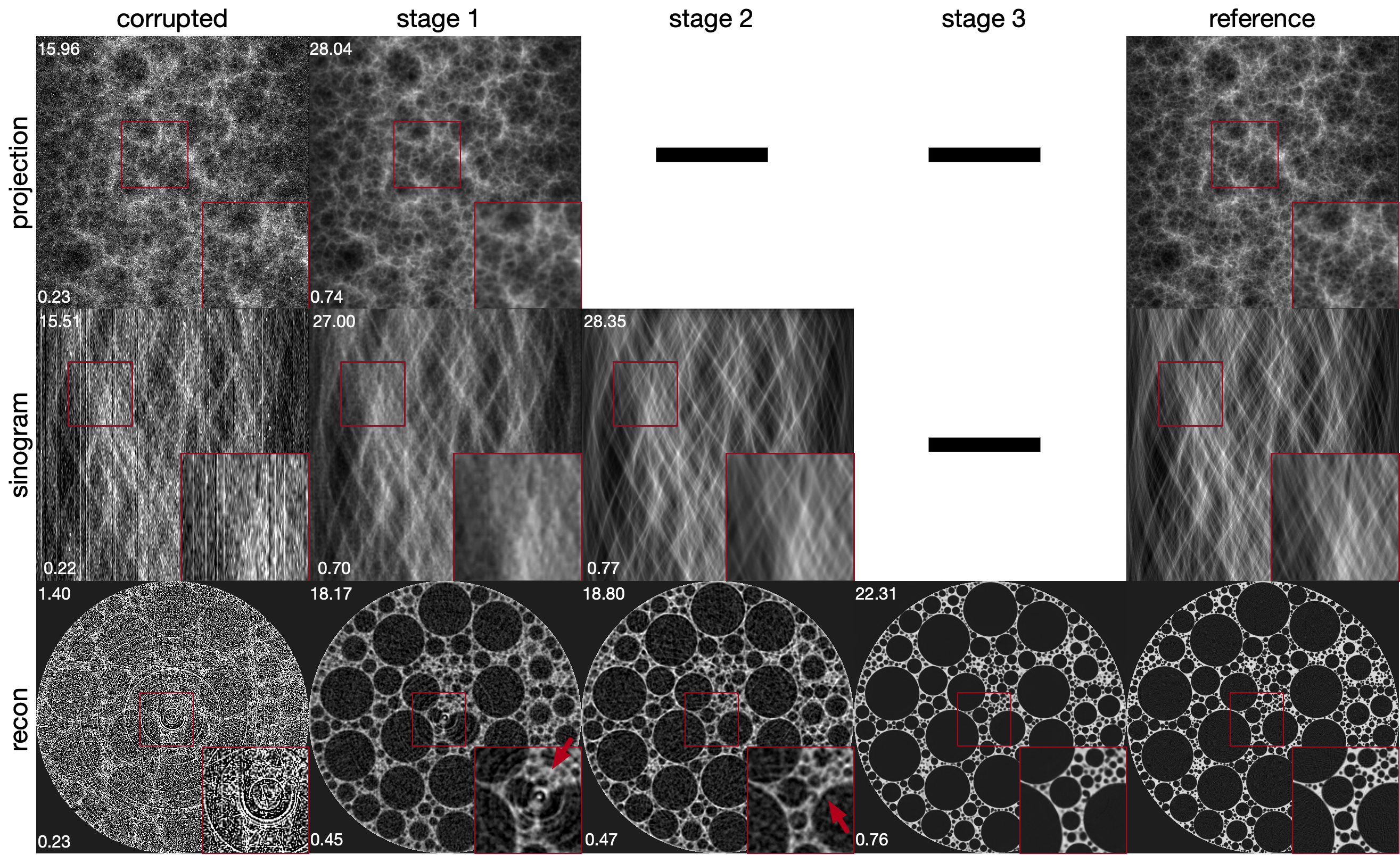}
\caption{Demonstration of artifact reduction across various processing stages. In the projection stage, zinger artifacts and portions of ring artifacts are mitigated. Subsequent processing in the sinogram stage eliminates remaining ring artifacts, while the reconstruction stage focuses on noise reduction to further improve image quality. For illustrative clarity, sinogram images are cropped into square formats.}
\label{fig:artifact_reduction_stages}
\end{figure}

\textbf{Artifact reduction in each stage:} 
The impact of our multi-stage strategy on artifact reduction was evaluated by examining results at each stage, as shown in Figure \ref{fig:artifact_reduction_stages}. Initial projection image processing effectively reduced most zinger artifacts and some ring artifacts, with PSNR and SSIM improving to 18.17 dB and 0.45, respectively. Subsequent sinogram processing reduced most ring artifacts, though at the expense of some high-resolution details, further enhancing PSNR to 18.80 dB and SSIM to 0.47. The final stage of processing the reconstructed images restored numerous image details, smoothed the image, and significantly elevated the PSNR to 22.25 dB and SSIM to 0.76.

\begin{figure}
\centering
\includegraphics[width=1\textwidth]{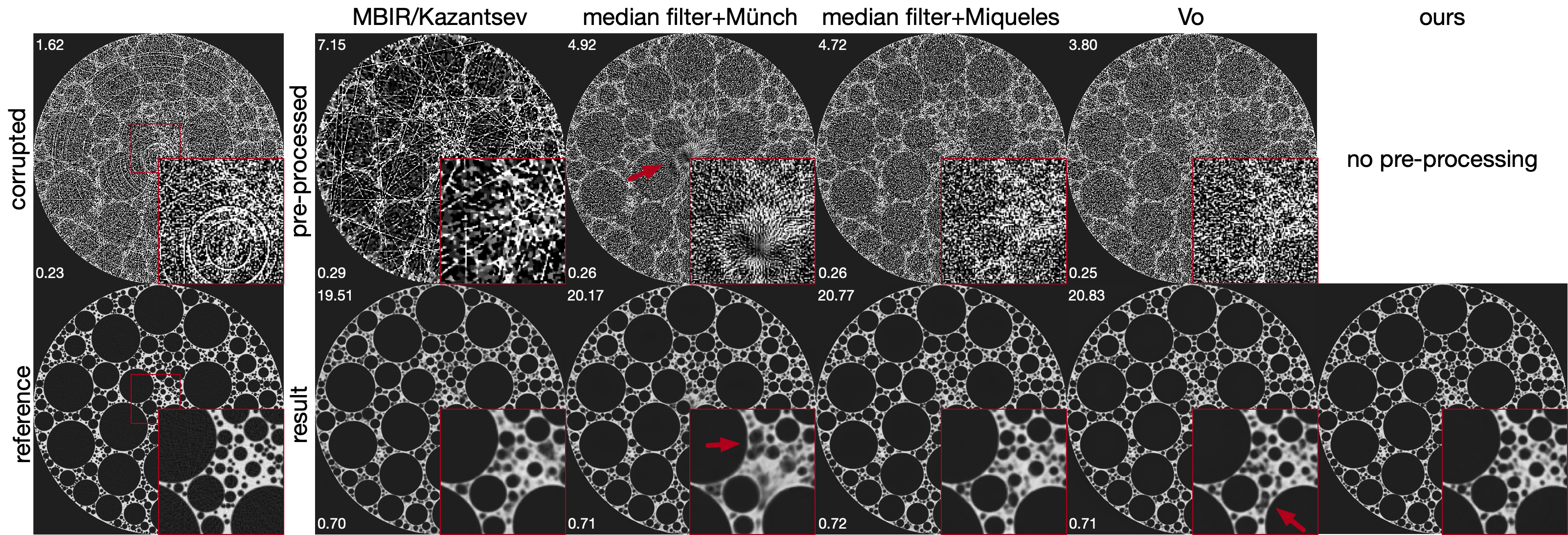}
\caption{Comparative visualization of artifact reduction efficacy using classical methods, where pre-processed reconstructed images serve as inputs for further post-processing. Despite the efficiency of classical methods in mitigating specific artifacts, the simultaneous presence of various global artifacts together with noise often results in inferior performance relative to our approach. Red insets highlight magnified sections for detailed examination. Quantitative assessments, indicated by PSNR and SSIM metrics, are displayed in the top-left and bottom-left corners of each image, respectively.}
\label{fig:compare_classical}
\end{figure}

\textbf{Classical methods:} 
After optimizing parameters on the training dataset, MBIR/Kazantsev \cite{kazantsev2017novel} demonstrated effective ring artifact reduction, as shown in Figure \ref{fig:compare_classical}. However, this method only partially removed zinger artifacts and caused oversmoothing in image details, resulting in some denoising effect. Although MBIR/Kazantsev \cite{kazantsev2017novel} produced images with the highest PSNR and SSIM values among all classical pre-processing methods (first row in Figure \ref{fig:compare_classical}), it resulted in lowest PSNR and SSIM when followed by deep learning post-processing (second row in Figure \ref{fig:compare_classical}). This indicates that successful ring artifact reduction by methods like \cite{kazantsev2017novel} does not necessarily improve final image detail and quality after post-processing. Techniques combining median filtering and wavelet-based ring reduction \cite{munch2009stripe} effectively addressed ring and zinger artifacts but introduced new artifacts, complicating post-processing. Median filtering in combination with Miqueles \cite{miqueles2014generalized} and Vo \cite{vo2018superior} managed to reduce artifacts without introducing new ones. Nonetheless, they were unable to deliver precise image details after post-processing. Contrarily, our method bypasses traditional artifact reduction steps, eliminating the need for manual parameter selection. Parameters are instead determined through a data-driven approach, yielding superior artifact reduction performance compared to classical methods followed by post-processing.
 
\subsection{Ablation Analysis}
\label{sec:ablation}
An ablation study was performed to assess the contribution and efficacy of the multi-stage strategy. Central to our design are the \textit{bypass connections}, aimed at enhancing artifact reduction and maintaining data integrity. Utilizing only processed data from previous stages might lead to the propagation of errors across the pipeline, with limited opportunity for rectification. By introducing both raw and processed data as inputs for subsequent stages, we aim to enrich the information available for following stages, potentially improving artifact reduction while preserving fidelity to the original dataset.

Our analysis included training models solely on projection, sinogram, and reconstruction images for artifact reduction and comparing the outcomes with those from our comprehensive multi-stage approach. Furthermore, we examined a variant of our multi-stage method that omitted \textit{bypass connections}, thereby directly linking each stage without incorporating raw inputs in subsequent phases (green flow in Figure \ref{fig:algorithm}). To ensure a fair comparison, single-stage models were adjusted to match the total trainable parameters of our multi-stage framework. As indicated in Table \ref{tab:abalation}, our multi-stage method consistently outperformed the alternatives, achieving significantly higher PSNR/SSIM values. This underscores the effectiveness of our multi-stage strategy. Notably, the inclusion of raw data as an additional input source yielded further enhancements in PSNR/SSIM values compared to the multi-stage method without raw data, reinforcing the value of integrating raw data for superior artifact reduction.

\begin{table}
\caption{Ablation study evaluating the impact of different stages of deep learning processing within our multi-stage method. This includes comparisons with scenarios processing only a single stage (projection, sinogram, or reconstruction) and a variant of our method that does not incorporate raw data but relies solely on the processed output from the previous stage. The assessments were performed on the simulated foam dataset, configured with $I_0, P_{\text{ring}}, P_{\text{zinger}}$ parameters set to $30, 0.1, 0.001$, respectively. The PSNR/SSIM values for the corrupted reconstructed images were $1.14$ and $0.23$.}
\label{tab:abalation}
  \centering
  \begin{center}
    \begin{tabular}{ccc} \toprule[0.1pt]
    stage of processing & result \\ \midrule[0.05pt]
        projection  & 17.74/0.44 \\ \midrule[0.05pt]
        sinogram  & 17.97/0.44 \\ \midrule[0.05pt]
        reconstruction & 19.55/0.70 \\ \midrule[0.05pt]
        multi-stage w/o \textit{bypass connections}  & 21.03/0.74\\ \midrule[0.05pt] 
        multi-stage w \textit{bypass connections} (\textbf{ours}) & \textbf{21.80}/\textbf{0.76}\\ 
    \bottomrule[0.1pt]
    \end{tabular}
\end{center}
\end{table}

\subsection{Computation Analysis}
A computational efficiency comparison was conducted between our method and classical methods followed by post-processing techniques using a simulated foam phantom dataset. This dataset comprised 256 corrupted projection images, each with dimensions of $512\times512$ pixels. All computational analyses were performed on a workstation equipped with an Intel i7-11700KF CPU and an Nvidia RTX3070 GPU.

Our approach processes the raw projections through three distinct stages, utilizing a separate CNN for each stage. Notably, in the sinogram stage, processed sinogram images underwent a fourfold upsampling. Training durations for the individual stages of our multi-stage approach were 9 minutes, 6 hours 50 minutes, and 3 hours 38 minutes, respectively, culminating in a total training period of 10 hours and 37 minutes. By comparison, the training time for the post-processing technique totaled approximately 14 hours and 31 minutes.

\begin{figure}
\centering
\includegraphics[width=0.6\textwidth]{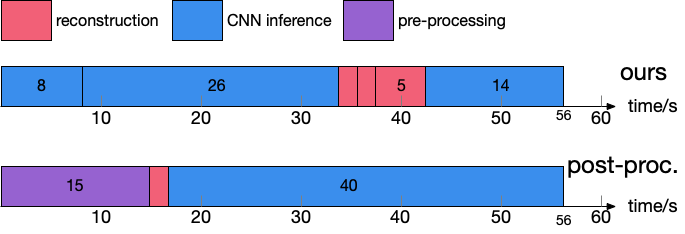}
\caption{Comparison of inference time between our method and classical method followed by post-processing. Our method involves reconstructing the raw projection, the processed projection, and the processed sinogram, requiring three passes through CNNs at each stage. In contrast, post-processing involves pre-processing the raw projection, reconstructing and applying it to the image domain CNN, exemplified here with a wavelet-based technique for clarity. Notably, when considering methods by Vo \cite{vo2018superior} and MBIR/Kazantsev \cite{kazantsev2017novel}, our approach demonstrates improved computational efficiency.}
\label{fig:inferencetime}
\end{figure}

As depicted in Figure \ref{fig:inferencetime}, the aggregate inference duration for our methodology was 56 seconds, identical to using wavelet-based method followed by post-processing as an example. The computation times required for the pre-processing stages of various classical artifact reduction methods are detailed in Table \ref{tab:compute_classical_method}. Among these, the method by MBIR/Kazantsev \cite{kazantsev2017novel}, which employs a regularized model-based iterative reconstruction for artifact reduction, was noted for its extended computation time, potentially hindering its practical application. In general, the computation time for our integrated multi-stage method aligns closely with the combined computational time of classical methods and their subsequent post-processing.


\begin{table}
\caption{Comparison of inference time between classical artifact reduction methods followed by post-processing on $512^3$ volumes with 256 projection angles.}
\label{tab:compute_classical_method}
\resizebox{\textwidth}{!}{
  \centering
    \begin{tabular}{c|c|c|c} \toprule[0.1pt]
    method & pre-processing & reconstruction & post-proc. \\ \midrule[0.05pt]
        median filter+M{\"u}nch \cite{munch2009stripe}  & 15s & \multirow{4}{*}{1s} & \multirow{5}{*}{40s} \\ \cmidrule{1-2}
        median filter+Miqueles \cite{miqueles2014generalized} & 15s &  \\ \cmidrule{1-2}
        Vo \cite{vo2018superior} & 31s & \\ \cmidrule{1-3}
        MBIR/Kazantsev \cite{kazantsev2017novel} & \multicolumn{2}{c|}{10mins}\\
    \bottomrule[0.1pt]
    \end{tabular}
}
\end{table}

\section{Discussion}
\label{section6}
The experiments performed in this paper indicate that the proposed multi-stage method effectively reduces artifacts in CT images, outperforming classical methods combined with post-processing and deep learning-based post-processing in PSNRs and SSIMs. Our method achieves accurate artifact reduction on both the simulated and experimental datasets. In particular, our method demonstrates a greater advantage over post-processing when severe ring and zinger artifacts are present in the reconstructed images. 

Our method has several advantages over existing methods. First, it employs a multi-stage approach that reduces artifacts accurately in their natural domain, where the artifact is easier to reduce than in other domains. By processing data in the projection, sinogram, and reconstruction domains, our method can effectively reduce different artifacts jointly. This multi-stage strategy also holds potential for addressing challenges in extended-field-of-view (FOV) CT, as well as in propagation-based phase-contrast imaging, where artifacts from phase retrieval could further be addressed in their respective domains. Second, the neural networks in each stage can be selected and trained independently to optimize performance for specific artifacts, enabling computationally efficient training. Third, our method can be easily integrated into existing CT pipelines without reducing the throughput of pipelines. The method is conceptually applicable to cone-beam CT, particularly for smaller cone angles where the geometry closely approximates parallel-beam CT. Overall, our method provides an efficient and effective solution for reducing artifacts in CT images.

Although high-quality reference data can be acquired in various manners as explained, such as scanning with high-dose CT, using more sophisticated reconstruction methods, or additional post-processing steps,  our proposed method still relies on these extra steps to obtain high-quality reference data. These steps may require additional time and effort in real-world scenarios. Therefore, it is worth exploring the integration of our multi-stage strategy with self-supervised methods, for example by applying Noise2Inverse~\cite{hendriksen2020noise2inverse} in a multi-stage manner. This could potentially reduce the reliance on high-quality reference data and improve the applicability of our method in real-world scenarios. 

In principle, the proposed method for artifact reduction could be extended to other computational imaging modalities with similar processing pipelines. In many settings, reconstructed images are computed through a series of processing steps in different domains, with artifacts in the measurements propagating through the pipeline resulting in artifacts in the reconstructed images. The key idea of applying deep learning within the pipeline instead of only at the end of the pipeline could be beneficial to improve the image quality of other imaging modalities, e.g. magnetic resonance imaging (MRI) or positron emission tomography (PET). By incorporating our multi-stage strategy into these modalities, artifact reduction can be achieved in their natural domain, potentially leading to improved image quality.

\section{Conclusion}
\label{section7}
In this work, we present a novel multi-stage artifact reduction method for CT images. Our approach involves three stages, each targeting a different type of image artifact in its corresponding domain: projection, sinogram, and reconstruction. We employ three separate neural networks, one for each stage, to jointly reduce artifacts in their respective domains. To reduce the risk of error propagation typical in conventional processing pipelines, we incorporate \textit{bypass connections} between stages. The networks are trained independently from each other in a sequential manner, ensuring computationally efficient training. Our experimental results demonstrate that our method outperforms deep learning-based post-processing techniques in terms of artifact reduction accuracy, achieving superior PSNRs and SSIMs both for simulated data and real-world experimental data. Moreover, our approach is designed to seamlessly integrate into existing CT pipelines for enhancing image quality.

\appendix

\section{Parameters Optimization of Classical Methods}

This appendix details the optimization of parameters for classical artifact reduction methods, employing median filtering and wavelet-based ring reduction \cite{munch2009stripe} as example on a simulated foam phantom dataset with settings $I_0 = 100$, $P_{\text{ring}} = 0.1$, and $P_{\text{zinger}} = 0.001$.

\begin{figure}
\centering
\includegraphics[width=0.9\textwidth]{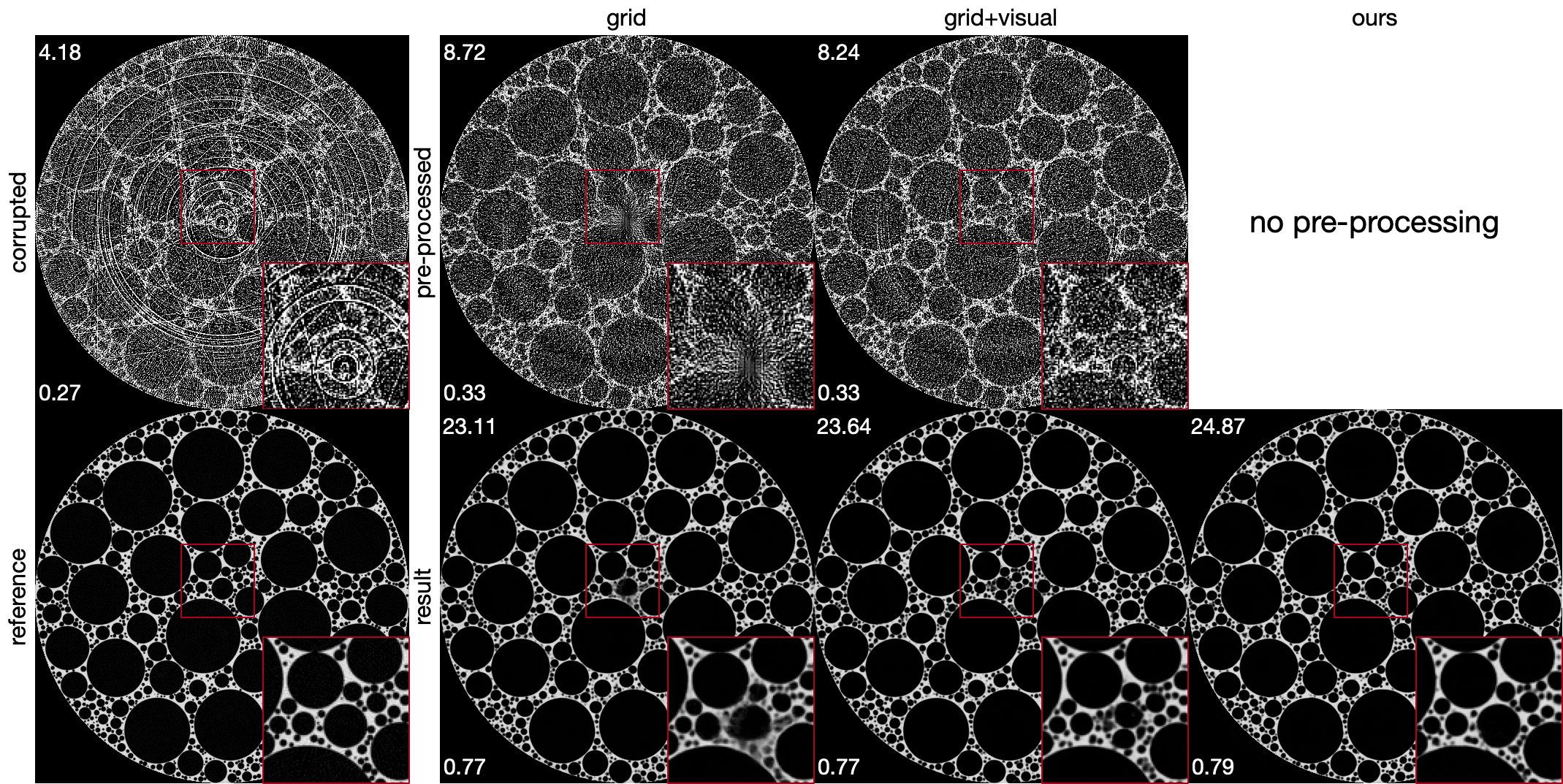}
\caption{Visual and quantitative comparison of artifact reduction outcomes using classical methods versus our proposed approach. Automatically selected parameters through grid search introduced unintended artifacts, complicating subsequent post-processing. Parameters refined through visual inspection offered better post-processing outcomes by compromising slightly on ring artifact reduction. Nonetheless, our method substantially surpasses these approaches in both visual quality and metric evaluations.}
\label{fig:comparewithclassical}
\end{figure}

Through grid search, a parameter set was identified that, while reducing artifacts, introduced new ones into the reconstructed images. A subsequent visual inspection allowed for the refinement of parameters, striking a balance between minimizing ring artifacts and avoiding the introduction of new ones. This tailored parameter set facilitated enhanced artifact reduction in subsequent post-processing steps. Despite these improvements, our proposed method significantly outperforms these classical approaches, as evidenced both visually in Figure \ref{fig:comparewithclassical} and quantitatively in Table \ref{tab:comparewithclassical}.

\begin{table}[5]
\setlength{\tabcolsep}{3pt}
\caption{Determined parameters of classical methods: (\romannumeral 1) the expected difference value between the outlier value and the median value of the image (dif), (\romannumeral 2) the median filter size (size) for outlier removal function, (\romannumeral 3) the discrete wavelet transform levels (level), (\romannumeral 4) the type of wavelet filter (type), and (\romannumeral 5) the damping parameter in Fourier space (sigma).}
\label{tab:classicalparameters}
  \centering
  \begin{center}
    \begin{tabular}{cccccc} \toprule[0.1pt]
     approach & (\romannumeral 1) dif & (\romannumeral 2) size & (\romannumeral 3) level & (\romannumeral 4) wname & (\romannumeral 5) sigma  \\ \midrule[0.05pt]
        grid & 0.5 & 3 & 4 & sym5 & 8 \\ 
        grid+visual & 0.5 & 3 & 4 & sym5 & 1\\ 
        \bottomrule[0.1pt]
    \end{tabular}
    \end{center}
\end{table}

\begin{table}[6]
\caption{Comparison of the average PSNR and SSIM values of our proposed method with classical methods combined with deep learning-based post-processing on the simulated foam phantom dataset. The values are calculated as the average of all slices.}
\label{tab:comparewithclassical}
  \centering
  \begin{center}
    \begin{tabular}{cccc} \toprule[0.1pt]
    & \multicolumn{2}{c}{PSNR/SSIM} \\
    method & pre-processed & result \\ \midrule[0.05pt]
        no artifact removal & 4.07/0.27 & - \\
        grid & 8.58/0.33 & 22.76/0.77 \\
        grid+visual & 8.10/0.33 & 23.41/0.77 \\
        our & - & \textbf{24.50}/\textbf{0.79}\\
    \bottomrule[0.1pt]
    \end{tabular}
    \end{center}
\end{table}



\ack{Acknowledgements}

     This research was co-financed by the European Union H2020-MSCA-ITN-2020 under grant agreement no. 956172 (xCTing). We acknowledge the use of the large language model, ChatGPT, to assist in refining the text. The tool was utilized at the sentence level for tasks such as correcting grammar and rephrasing sentences.

\referencelist[Multi-stageArtifactReduction]

\end{document}